\documentclass{ws-p9-75x6-50}
\usepackage{epsfig}

\def\eg{e.g.\hbox{}}
\def\etal{{\it et~al.\/}}

\def\Cplusplus{\hbox{C\raise.20ex\hbox{\footnotesize ++}}}
\def\constant{{\rm constant}}

\begin{document}
\title{A Multiple-Grid-Patch Evolution Scheme for 3-D Black Hole Excision}
\author{J.~Thornburg}
\address{Institut f\"{u}r Theoretische Physik, Universit\"{a}t Wien\\
	 Boltzmangasse 5, A-1090 Wien, Austria\\
   E-mail: jthorn@thp.univie.ac.at}


%
%

\maketitle
\abstracts{
When using black hole excision to numerically evolve a fully generic
black hole spacetime, most 3-D $3+1$ codes use an $xyz$-topology (spatial)
grid.  In such a grid, an $r = \constant$ excision surface must be approximated
by an irregular and non-smooth "staircase-shaped" excision grid boundary,
which may introduce numerical instabilities into the evolution.
In this paper I describe an alternate scheme, which uses multiple grid
patches, each with topology $\{r \times ({\rm angular\ coordinates})\}$,
to cover the slice outside the $r = \constant$ excision surface.  The
excision grid boundary is now smooth, so the evolution should be less
prone to instabilities.  With 4th~order finite differencing, this
code evolves Kerr initial data to ${\sim}\, 60M$ using the ADM equations;
I'm currently implementing the BSSN equations in it in the hope that this
will improve the stability.}


\section{Introduction}

Black hole excision (BHE)
\cite{Thornburg-1987-2BH-initial-data,Seidel-Suen-1992-BHE}
is now widely used for the $3+1$ numerical evolution of dynamic
black hole spacetimes.  In 3-D, most codes use $xyz$-topology (spatial)
grids to avoid $z$~axis problems, with a ``staircase-shaped'' excision
surface approximating $r=\constant$.  Unfortunately, the irregular
shape of such an inner boundary to the grid considerably complicates
the finite differencing, and often introduces numerical instabilities
(\eg{} \cite{2BH-grand-challenge-alliance-1998-moving-BH}).


\section{Multiple Grid Patches}

To avoid the numerical problems caused by irregular grid boundaries,
I propose a new scheme, in which multiple grid patches, each with
topology $\{r \times ({\rm angular\ coordinates})\}$, are used to
cover the slice outside the $r = \constant$ excision surface.  This
allows the inner boundary to be smooth.  By using multiple patches,
the angular coordinates can be non-singular in a neighborhood of each
patch, yet still cover the full angular domain.

\begin{figure}[b]
\vspace{-20mm}
\begin{center}
\epsfig{file=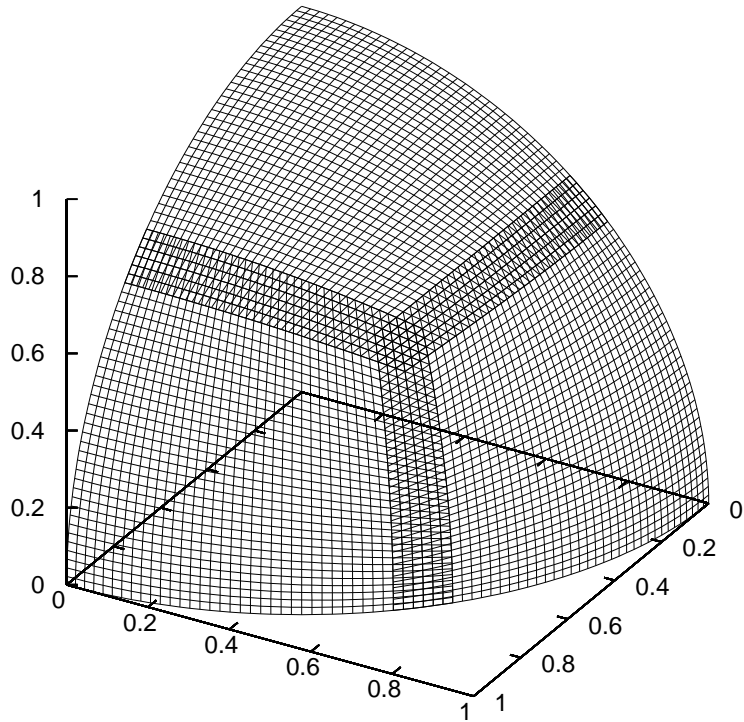,width=120mm}
\end{center}
\vspace{-10mm}
\caption[Example of 3-patch grid system]
	{
	This figure shows an $r = \constant$ shell of a
	3-patch coordinate/grid system covering one octant
	of space outside the excision surface $r = r_{\min}$.
	The patches overlap by 3~grid spacings.
	}
\label{fig-octant-3patch-system}
\end{figure}

Figure~\ref{fig-octant-3patch-system} shows an example of a
3-patch coordinate/grid system of this type, covering one octant
of space.  There are many possible ways of constructing such a multi-patch
coordinate/grid system; the particular choice used here has the advantage
that each pair of adjacent patches shares both $r$ and and one of the
angular coordinates.  Each patch is surrounded by angular ``ghost zones''
(not shown in the figure); field variables in the ghost zones are
interpolated from neighboring patches.  Because of the shared coordinates
between adjacent patches, only 1-dimensional (tensor) interpolation
is needed.  By overlapping the patches a few grid spacings, the
interpolations always use data from patch {\em interiors\/}, which
helps the evolution's numerical stability.


\section{Numerical Results}

I have implemented this scheme in a new BHE evolution code
(about 35K lines of Maple and \Cplusplus), using 4th~order finite
differencing in space, a smoothly-nonuniform radial grid, and a
4th~order Runge-Kutta method of lines time integration, all basically
as in my spherically symmetric code (\cite{Thornburg-1999-sssf-evolution}).
With the standard ADM equations, the code currently has finite differencing
instabilities at the corners where all 3~patches meet, with the energy
constraint reaching $\sim 1\%$ by $t \sim 60M$ for Kerr initial data
with spin~0.6.  I'm currently implementing the BSSN formulation
\cite{Alcubierre-etal-2000-BSSN-evolution} and investigating changes
to the interpolation scheme to try to improve the stability.


%
%


\end{document}